# Enhanced dynamic homogenization of hexagonally packed granular materials with elastic interfaces

Andrea Bacigalupo[1] and Luigi Gambarotta[1*]

[1]Department of Civil, Chemical and Environmental Engineering, University of Genova, Italy

**Abstract**

It is well known that the classical energetically consistent micropolar model has limits in simulating the frequency band structure of packed granular materials (see Merkel *et al*., 2011). It is here shown that if a standard continualization of the difference equation of motion of the discrete model is carried out, an equivalent micropolar continuum is obtained which is able to accurately simulate the optical branches of the discrete model. Nevertheless, this homogenized continuum presents non-positive defined elastic potential energy, a deficiency that limits the reliability of the model and implies instability phenomena (destabilizing effects) in the acoustic branches. This drawback is circumvented here through an high-frequency dynamic homogenization scheme which is based on an enhanced continualization of the discrete governing equations into pseudo-differential equations. Through a formal Taylor expansion of the pseudo-differential operators a higher order differential equation corresponding to the governing equation of a non-local continuum thermodynamically consistent are obtained. The resulting approach allows obtaining an equivalent micropolar continuum characterized by inertial non-locality. Moreover, higher order continua with non-local constitutive and inertial terms may be derived. The proposed continuum models are proved to be able to accurately describe both the static and dynamic behavior of the discrete granular model. Finally, the convergence to the response of the discrete system is shown when increasing the order of the higher order continuum.

**Keywords:** Periodic materials; Metamaterials; Rigid block assemblages; Elastic interfaces; Dispersive waves; Band gaps; Micropolar modelling.

___________________________________________________

[*] Corresponding Author

# 1. Introduction

Mechanical modeling of granular materials is a long-investigated topic of great theoretical and applicative interest. Micromechanical approaches based on the description of the ensemble of finite-sized particles, in most cases idealized as rigid spheres, are still being investigated by many Authors and the acquired knowledge proved useful for the description of the different static and dynamic behaviors of these materials by means of continuous models derived through homogenization techniques.

Several remarkable contributions to the formulation of micromechanically based continuum models have been given by Kanatani (1979), Christoffersen *et al.* (1981), Chang and Liao (1990), Chang and Ma (1991) among the others. The treatment of the roto-translational motion of the granules in the description of the mutual interactions naturally involves the micro-rotation as an independent kinematic variable and the consequent formulation of continuous micropolar models. Further contributions to the definition of the structure of the overall constitutive equations and of the properties of the stress state in the derived micropolar models were given by Bardet and Vardoulakis (2001), Jenkins and La Ragione (2001), Kruyt (2003), Kruyt and Rothenburg (2004). A micro-mechanically based approach for describing the large strain behavior of granular materials at the macro-scale was proposed by Miehe and Dettmar (2004) in which the micro-macro transition involves a periodic cell aggregate of discrete solid granules. Wellmann *et al.* (2008) developed a homogenization strategy for granular materials by averaging over periodic representative volume elements the microscopic quantities resulting from a DEM simulation of the rigid granules assembly including the effects of elasto-frictional contacts. Liu *et al.* (2019) proposed a multi-scale model for the analysis of granular systems, which combines the principles of a coupled FEM–DEM approach including a procedure for the definition of appropriate micro-scale boundary conditions.

Starting from a one-dimensional discrete model of spherical particle consisting in a Fermi-Pasta-Ulam oscillator, Mulhaus and Oka (1996) derived the governing equation of an equivalent continuum by replacing the difference expressions of the discrete model through appropriate differential expressions. This continuum turned out to be a combination of a micropolar and a strain gradient model and the Authors noted that such continuum was characterized by *destabilizing effects* in the higher order terms, i.e. loss of positive definiteness of the elastic potential density. The approach was extended to obtain a three-dimensional model for the granular material that shows dispersive propagation of elastic waves.

Suiker *et al.* (2000, 2001a, 2001b) described the granular material as a conglomerate of particles whose interaction is localized at representative contact points. The particle kinematics is governed by particle translation and rotation. The constitutive relations at the contacts relate relative displacements and rotation between particles in contact to the contact forces and couples. The generalized displacement at a particle is related to the macro-deformation gradients of the equivalent continuum through polynomial expansions and a second-gradient micro-polar continuum is derived, which exhibits some *destabilizing effects*. Acoustic branches are obtained showing a frequency maximum with a frequency decrease up to take vanishing values, while the optical branch, which turns out to be related to the micro-rotation of the granules, is increasing with the wave number. Suiker and de Borst (2005) extended the previous results concerning the derivation of continuum models to simulate the inhomogeneous mechanical behavior of granular assemblies. Strain gradient and strain-gradient micropolar models were derived and the wave dispersion curves were compared to corresponding discrete lattice model discussing the validity limits of the continuum model.

Pavlov *et al.* (2006) considered a granular material made up of square pattern of equal rigid circles undergoing translations and rotations with mutual elastic point contacts. The discrete Lagrangian model has been homogenized by approximating the generalized displacements of the granules appearing in the elastic potential energy of the discrete model through a Taylor expansion of a macro-displacement field. The wave propagation in the equivalent micropolar continuum has been analyzed. An analogous approach has been developed by Pasternak and Mühlhaus (2005) who considered a 1D chain of spherical rigid grains connected by shear and rotational springs and analyzed different homogenization procedures and studied the propagation of harmonic waves. The homogenization procedure based on the differential expansion of the generalized displacement field has been extended by Pasternak and Dyskin (2010, 2014). The elastic potential energy of particulate two-dimensional materials with periodic cells consisting in both square and hexagonal packing of cylindrical rigid bodies connected through elastic contact points have been approximated through a Taylor expansion of a macro-displacement field to obtain an equivalent micropolar continuum and the long wavelength dispersive propagation in the equivalent granular material were analyzed.

It is worth to note that theoretical (Merkel *et al.*, 2010) and experimental results (Merkel *et al.*, 2011) from a study focused on the influence of the rotational degree of freedom of the granules have highlighted some limitations of the micropolar model in simulating the dispersive wave propagation in periodic granular materials which were not considered in the previous literature. The criticism concerned the inability of the micropolar model to simulate the optical band structure of hexagonally packed granular materials in the neighborhood of long wavelengths. In fact, while the optical branch in the Bloch spectrum of the micropolar model is always characterized by an upward concavity, on the contrary the discrete granular model is characterized by a downward concavity of the optical branch. To circumvent such drawback an enhanced higher order micropolar equivalent continuum was proposed by Merkel and Luding (2017). Bacigalupo and Gambarotta (2017) have shown, in a more general context concerning periodic blocky materials with elastic interfaces, that it is possible to derive a micropolar equivalent continuum capable of simulating the original Lagrangian system. This can be done simply by applying a standard continualization, i.e. a differential expansion, in the downscaling law. Despite this encouraging result, it was shown that the homogenization based on the standard continualization necessarily implies the loss of positive definiteness of the elastic potential energy density of the equivalent continuum, even for higher order models. This outcome appears to be thermodynamically inconsistent. To solve this kind of problems, the Authors have recently proposed an enhanced dynamic homogenization procedure for lattice-like materials based on a transformation of the difference equation of motion of the discrete system into a pseudo-differential problem. The Taylor expansion of the pseudo-differential equations provides a higher order differential problem which is thermodynamically consistent and able to catch the Bloch spectra in the neighborhood of the long wavelengths (Bacigalupo and Gambarotta, 2019). This approach has been extended to multi-dimensional systems (Bacigalupo and Gambarotta, 2021) and is here applied to a simple hexagonal pattern of hexagonal rigid circles endowed with translational and rotational degrees of freedom. In Section 2 the discrete model is presented together with the equations governing the harmonic wave propagation. In Section 3 the micropolar equivalent continuum obtained by standard continualization is derived and analysed to show its validity limits. In Section 4 the differential equations of the enhanced micropolar continuum are derived together with the constitutive equations. A comparison of the Bloch spectra of the discrete granular model with the proposed micropolar model is carried out. The capabilities and the validity limits of the proposed model are highlighted.

## 2. Lagrangian model for the hexagonal packed granular material

Let consider a two-dimensional packed granular material made of rigid circles of radius $R$ periodically arranged in hexagonal pattern with coordination number $n = 6$. A reference circle is considered together with the surrounding ones as shown in Figure 1. The block interaction is assumed to take place at a small linear elastic interface of length $b$ localized around the contact point between connected circle with normal $K_n$, tangential $K_t$ and rotational $K_\varphi = \frac{1}{12} K_n b^2$ linear elastic stiffnesses, respectively. Each circle has mass $M$ and mass moment of inertia $J = \frac{1}{2} MR^2$.

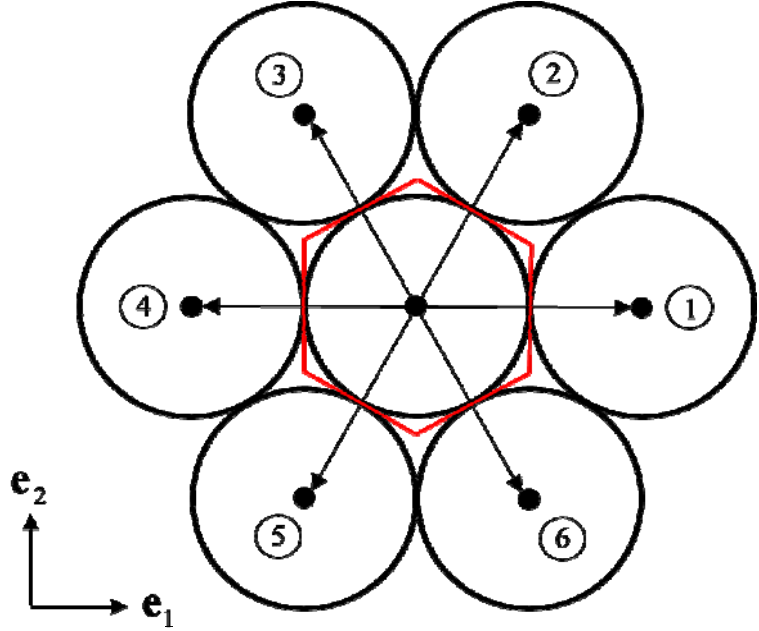

Figure 1. Hexagonal packed granular material.

Let denote with $\mathbf{d} = \{u \quad v \quad \varphi\}^T$ the generalized displacement vector of the reference circle, located at the centre of the periodic cell, and with $\mathbf{d}_i = \{u_i \quad v_i \quad \varphi_i\}^T$ the generalized displacement of the $i$-th surrounding circle (with $i = 1,..,6$). Moreover let assume that the resultant force and resultant moment acting on the centre of the reference circle are represented by vector $\mathbf{f} = \{f_u \quad f_v \quad c\}^T$. Once assumed the following ratios $r = \dfrac{b}{R}$ and $r_1 = \dfrac{K_t}{K_n}$, the equation of motion of the reference circle takes the following form (see for instance Bacigalupo and Gambarotta, 2017)

$$-K_n \left( \begin{array}{l} -u_4 + 3(r_1+1)u - u_1 - \dfrac{1}{4}(3r_1+1)(u_2 + u_3 + u_5 + u_6) + \\ + \dfrac{\sqrt{3}}{4}(r_1-1)(v_2 - v_3 + v_5 - v_6) - \dfrac{\sqrt{3}}{2}Rr_1(\phi_2 + \phi_3 - \phi_5 - \phi_6) \end{array} \right) + f_u - M\ddot{u} = 0,$$

$$-K_n \left( \begin{array}{l} \dfrac{\sqrt{3}}{4}(r_1-1)(u_2 - u_3 + u_5 - u_6) - r_1 v_1 + 3(r_1+1)v - r_1 v_4 + \\ -\dfrac{1}{4}(3+r_1)(v_2 + v_3 + v_5 + v_6) + \\ +\dfrac{1}{2}r_1 R(2\phi_1 + \phi_2 - \phi_3 - 2\phi_4 - \phi_5 + \phi_6) \end{array} \right) + f_v - M\ddot{v} = 0, \quad (1)$$

$$-K_n \left\{ \begin{array}{l} \dfrac{\sqrt{3}}{2}Rr_1(u_2 + u_3 - u_5 - u_6) - \dfrac{1}{2}Rr_1(2v_1 + v_2 - v_3 - 2v_4 - v_5 + v_6) + \\ -\dfrac{1}{12}R^2(r^2 - 12r_1)(\phi_1 + \phi_2 + \phi_3 + \phi_4 + \phi_5 + \phi_6) + \dfrac{1}{2}R^2(r^2 + 12r_1)\phi \end{array} \right\} + c - J\ddot{\phi} = 0.$$

Two-dimensional harmonic wave propagation in such periodic material along the generic axis identified by the unit vector $\mathbf{i}$ is analyzed by assuming the motion of the $i$-th circle in the form $\mathbf{d}_i = \hat{\mathbf{d}} \exp\left[I(\mathbf{k} \cdot \mathbf{x}_i - \omega t)\right]$, being $\hat{\mathbf{d}} = \{\hat{u} \quad \hat{v} \quad \hat{\phi}\}^T$ the polarization vector, $\mathbf{k} = q\mathbf{i}$ the wave vector, $q$ and $\omega$ the wave number and the circular frequency, respectively, and $\mathbf{x}_i$ the position vector of the $i$-th circle with respect to the reference circle. Under this assumption and by ignoring the applied generalized force $\mathbf{f}$, the equation of motion of the reference circle takes the form of an eigenvalue problem

$$\left( \begin{bmatrix} A_{uu} & A_{uv} & A_{u\phi} \\ A_{vu} & A_{vv} & A_{v\phi} \\ A_{\phi u} & A_{\phi v} & A_{\phi\phi} \end{bmatrix} - \omega^2 \begin{bmatrix} M & 0 & 0 \\ 0 & M & 0 \\ 0 & 0 & J \end{bmatrix} \right) \hat{\mathbf{d}} = \mathbf{0}, \quad (2)$$

where the element of the Hermitian matrix are expressed in terms of the component $k_1$ and $k_2$ of the wave vector

$$\begin{aligned}
A_{uu} &= K_n\left[3(r_1+1)-(3r_1+1)\cos(k_1 R)\cos(\sqrt{3}k_2 R)-2\cos(2k_1 R)\right], \\
A_{uv} &= A_{vu} = -\sqrt{3}(r_1-1)K_n\sin(k_1 R)\sin(\sqrt{3}k_2 R), \\
A_{u\phi} &= -A_{\phi u} = -2\sqrt{3}I\, r_1 K_n R\cos(k_1 R)\sin(\sqrt{3}k_2 R), \\
A_{vv} &= K_n\left[3(1+r_1)-(3+r_1)\cos(k_1 R)\cos(\sqrt{3}k_2 R)-2r_1\cos(2k_1 R)\right], \\
A_{v\phi} &= -A_{\phi v} = 2I\, K_n R r_1\left[\sin(k_1 R)\cos(\sqrt{3}k_2 R)+\sin(2k_1 R)\right], \\
A_{\phi\phi} &= K_n R^2\left\{\tfrac{1}{2}(r^2+12r_1)-\tfrac{1}{6}(r^2-12r_1)\cos(2k_1 R)-\tfrac{1}{3}(r^2-12r_1)\cos(k_1 R)\cos(\sqrt{3}k_2 R)\right\},
\end{aligned} \qquad (3)$$

being invariant to $\pi/3$ rotations (hexagonal symmetry).

The eigenvalue problem (2) is solved by three dispersion functions $\omega_i(\mathbf{k})$ (with $i=1,3$) in the Bloch plane. Two solutions represent acoustic branches, with vanishing frequency at the long wavelength limit $|\mathbf{k}|\to 0$, the third one is an optical branch departing from a critical point with vanishing group velocity and angular frequency $\omega_{opt}(|\mathbf{k}|=0)=\sqrt{\dfrac{A_{\phi\phi}(k_1=k_2=0)}{J}}=\sqrt{\dfrac{12 r_1 K_n R^2}{J}}=\sqrt{24 r_1 \dfrac{K_n}{M}}=\sqrt{24\dfrac{K_t}{M}}$ that only depends on the tangential stiffness and the mass circle. It may be easily derived that in the neighborhood of the critical point the optical dispersion function is convex if $(r^2-12r_1)>0$. But this circumstance appears to be rather unlikely in consideration of the smallness of the ratio $r=\dfrac{b}{R}$ for the assumed material microstructure. As a consequence the optical dispersion surface is concave in the neighbourhood of the critical point in the Bloch space, an outcome already obtained by Merkel *et al.* (2011).

## 3. Micropolar continuum via standard continualization

To derive an equivalent continuum model from the discrete Lagrangian model considered in Section 2 a standard approach is usually assumed in literature (see Bacigalupo and Gambarotta, 2019) which is based on the approximation of the generalized displacements of the surrounding circles through a Taylor expansion of a continuum macro displacement field $\mathbf{d}_M(\mathbf{x}) = \{U(\mathbf{x}) \quad V(\mathbf{x}) \quad \Phi(\mathbf{x})\}^T$. Following this approach, called standard continualization, the generalized displacements of the *i*-th node is approximated through a second order expansion of the continuum macro-displacement fields

$$\begin{aligned}
u_i(t) &\cong U(\mathbf{x},t) + 2R\,\nabla U(\mathbf{x},t)\mathbf{n}_i + 2R^2\,(\nabla\otimes\nabla)U(\mathbf{x},t):(\mathbf{n}_i\otimes\mathbf{n}_i), \\
v_i(t) &\cong V(\mathbf{x},t) + 2R\,\nabla V(\mathbf{x},t)\mathbf{n}_i + 2R^2\,(\nabla\otimes\nabla)V(\mathbf{x},t):(\mathbf{n}_i\otimes\mathbf{n}_i), \\
\phi_i(t) &\cong \Phi(\mathbf{x},t) + 2R\,\nabla\Phi(\mathbf{x},t)\mathbf{n}_i + 2R^2\,(\nabla\otimes\nabla)\Phi(\mathbf{x},t):(\mathbf{n}_i\otimes\mathbf{n}_i),
\end{aligned} \quad (4)$$

involving the macro-displacement gradient and second gradient, where $\chi = \nabla\Phi$ is the macro-curvature curvature. By substituting the approximations (4) in the discrete equations of motion (1), the equation of motion of a micropolar equivalent continuum are obtained

$$\begin{cases}
\dfrac{3}{2}K_n R^2\left[(3+r_1)\dfrac{\partial^2 U}{\partial x_1^2} + (1+3r_1)\dfrac{\partial^2 U}{\partial x_2^2}\right] + 3(1-r_1)K_n R^2\dfrac{\partial^2 V}{\partial x_1 \partial x_2} + 6r_1 K_n R^2\dfrac{\partial \Phi}{\partial x_2} - M\ddot{U} = 0, \\
3(1-r_1)K_n R^2\dfrac{\partial^2 U}{\partial x_1 \partial x_2} + \dfrac{3}{2}K_n R^2\left[(1+3r_1)\dfrac{\partial^2 V}{\partial x_1^2} + (3+r_1)\dfrac{\partial^2 V}{\partial x_2^2}\right] - 6r_1 K_n R^2\dfrac{\partial \Phi}{\partial x_1} - M\ddot{V} = 0, \\
-6r_1 K_n R^2 \dfrac{\partial U}{\partial x_2} + 6r_1 K_n R^2 \dfrac{\partial V}{\partial x_1} - 12r_1 K_n R^2 \Phi + \dfrac{1}{2}K_n R^4(r^2 - 12r_1)\Delta\Phi - J\ddot{\Phi} = 0.
\end{cases} \quad (5)$$

Denoting with $\rho_m$ the mass density of the material inside the circles and with $\rho$ averaged density of the periodic cell of area $A_{hex}$, it follows $M = \pi\rho_m R^2 = \rho A_{hex} = 4\sqrt{3}\rho R^2$, from which one obtains $\rho = \dfrac{\pi\sqrt{3}}{12}\rho_m$ and $J = \dfrac{\pi}{2}\rho_m R^4 = 2\sqrt{3}\rho R^4$. Accordingly, equations (5) may be rewritten in the form

$$\begin{cases}
\dfrac{\sqrt{3}}{8}(3+r_1)K_n\dfrac{\partial^2 U}{\partial x_1^2} + \dfrac{\sqrt{3}}{8}(1+3r_1)K_n\dfrac{\partial^2 U}{\partial x_2^2} + \dfrac{\sqrt{3}}{4}(1-r_1)K_n\dfrac{\partial^2 V}{\partial x_1 \partial x_2} + \dfrac{\sqrt{3}}{2}r_1 K_n\dfrac{\partial \Phi}{\partial x_2} - \rho\ddot{U} = 0, \\
\dfrac{\sqrt{3}}{4}(1-r_1)K_n\dfrac{\partial^2 U}{\partial x_1 \partial x_2} + \dfrac{\sqrt{3}}{8}(1+3r_1)K_n\dfrac{\partial^2 V}{\partial x_1^2} + \dfrac{\sqrt{3}}{8}(3+r_1)K_n\dfrac{\partial^2 V}{\partial x_2^2} - \dfrac{\sqrt{3}}{2}r_1 K_n\dfrac{\partial \Phi}{\partial x_1} - \rho\ddot{V} = 0, \\
+\dfrac{\sqrt{3}}{24}K_n R^2(r^2 - 12r_1)\Delta\Phi - \sqrt{3}r_1 K_n \Phi + \dfrac{\sqrt{3}}{2}r_1 K_n\dfrac{\partial V}{\partial x_1} - \dfrac{\sqrt{3}}{2}r_1 K_n\dfrac{\partial U}{\partial x_2} - \dfrac{1}{2}\rho R^2\ddot{\Phi} = 0.
\end{cases} \quad (6)$$

By comparing equation (6) with the canonical form of the equation of motion of a micropolar two-dimensional continuum with hexagonal symmetry (21) (see for instance Eremeyev *et al.*, 2013; Nowacki, 1986; Wang and Stronge, 1999) given in Appendix A, the four micropolar elastic moduli may be identified $\lambda = \frac{\sqrt{3}}{8}(1-r_1)K_n$, $\mu = \frac{\sqrt{3}}{8}(1+r_1)K_n$, $\kappa = \frac{\sqrt{3}}{4}r_1 K_n$, $\gamma = \frac{\sqrt{3}}{24}(r^2 - 12r_1)K_n R^2$, respectively, which have to satisfy the following conditions for the positive definiteness of the elastic potential energy density

$$\lambda + \mu = \frac{\sqrt{3}}{8}K_n > 0, \quad \mu = \frac{\sqrt{3}}{8}K_n(1+r_1) > 0, \quad \kappa = \frac{\sqrt{3}}{4}r_1 K_n > 0 \quad \gamma = \frac{\sqrt{3}}{24}K_n R^2 (r^2 - 12r_1) > 0.$$

While the first three conditions are unconditionally satisfied, the fourth one turns out to be satisfied if $\left(\frac{b}{R}\right)^2 - 12\frac{K_t}{K_n} > 0$, a circumstance that does not seems possible as already discussed in the previous Section. As a consequence, the micropolar equivalent continuum derived through the standard continualization does not results to be energetically consistent because the non-positivity of the elastic constant $\gamma$ relating the curvatures to micro-couples. In the limit case of point contact, i.e. $b \to 0$ and vanishing rotational stiffness $K_\varphi \to 0$, one obtains $\gamma < 0$, in agreement with Sulem and Mühlhaus (1997), equation (10). It may be easily shown that such pathology persists also when considering higher order Taylor expansions in approximating the nodal displacement in the equation of motion of the Lagrangian system.

The constitutive equation of the micropolar continuum with hexagonal symmetry (22) specialize

$$\begin{Bmatrix} \sigma_{11} \\ \sigma_{22} \\ \sigma_{12} \\ \sigma_{21} \\ m_1 \\ m_2 \end{Bmatrix} = \frac{\sqrt{3}}{8}K_n \begin{bmatrix} (3+r_1) & (1-r_1) & 0 & 0 & 0 & 0 \\ (1-r_1) & (3+r_1) & 0 & 0 & 0 & 0 \\ 0 & 0 & (1+3r_1) & (1-r_1) & 0 & 0 \\ 0 & 0 & (1-r_1) & (1+3r_1) & 0 & 0 \\ 0 & 0 & 0 & 0 & \frac{1}{3}R^2(r^2-12r_1) & 0 \\ 0 & 0 & 0 & 0 & 0 & \frac{1}{3}R^2(r^2-12r_1) \end{bmatrix} \begin{Bmatrix} \gamma_{11} \\ \gamma_{22} \\ \gamma_{12} \\ \gamma_{21} \\ \chi_1 \\ \chi_2 \end{Bmatrix}.$$

(7)

Moreover, it may be shown that the same homogenized model may be obtained through a variational derivation of the micropolar model based on the approximation (4) which is related to the Hill-Mandel macro homogeneity condition (see Bacigalupo and Gambarotta, 2017, for a general discussion concerning periodic blocky materials). Following this approach and retaining expansion (4) to the first order, a positive defined elastic modulus is obtained

$$\gamma^+ = \frac{\sqrt{3}}{24} K_n R^2 \left( r^2 + 12 r_1 \right).$$

The harmonic wave propagation in the micropolar two dimensional continua with hexagonal symmetry (23) is specialized in the form

$$\begin{bmatrix} \frac{\sqrt{3}}{8}\begin{bmatrix}(3+r_1)k_1^2+\\+(1+3r_1)k_2^2\end{bmatrix}-\frac{\rho}{K_n}\omega^2 & \frac{\sqrt{3}}{4}(1-r_1)k_1 k_2 & -I\frac{\sqrt{3}}{2}r_1 k_2 \\ \frac{\sqrt{3}}{4}(1-r_1)k_1 k_2 & \frac{\sqrt{3}}{8}\begin{bmatrix}(1+3r_1)k_1^2+\\+(3+r_1)k_2^2\end{bmatrix}-\frac{\rho}{K_n}\omega^2 & I\frac{\sqrt{3}}{2}r_1 k_1 \\ I\frac{\sqrt{3}}{2}r_1 k_2 & -I\frac{\sqrt{3}}{2}r_1 k_1 & \frac{\sqrt{3}}{24}R^2(r^2-12r_1)(k_1^2+k_2^2)+\\ & & +\sqrt{3}r_1-\frac{1}{2}\frac{\rho}{K_n}R^2\omega^2 \end{bmatrix} \begin{Bmatrix}\hat{U}\\ \hat{V}\\ \hat{\Phi}\end{Bmatrix} = \mathbf{0}.$$

(8)

The Hermitian matrix in the eigenvalue problem (8) may be shown to be a second order approximation of the Hermitian matrix of the eigenvalue problem (2) in the components of the wave vector (Bacigalupo and Gambarotta, 2017).

To evaluate the accuracy of the resulting continuum model, a comparison between the dispersion functions of the Lagrangian model plotted along the boundary of the non-dimensional irreducible first Brillouin zone (Brillouin, 1953) with those from the micropolar model derived from standard continualization is given in the diagrams of Figure 2 and 3 for different values of the parameter $r_1$. This Figures also include the diagrams from the energetically consistent micropolar model having positive define modulus $\gamma^+$. While in Figures 2.a and 3.a the non-dimensional angular frequency $\omega\sqrt{\frac{M}{K_n}}$ is plotted as function of the arch length $\Xi$ measured on the closed polygonal curve $\Upsilon$ with vertices identified by the values $\Xi_j$, $j = 0,1,2$, in Figures 2.b and 3.b the diagrams are plotted along the boundary of a sub-domain of the irreducible first Brillouin zone, which is homothetically scaled by a factor 0.5.

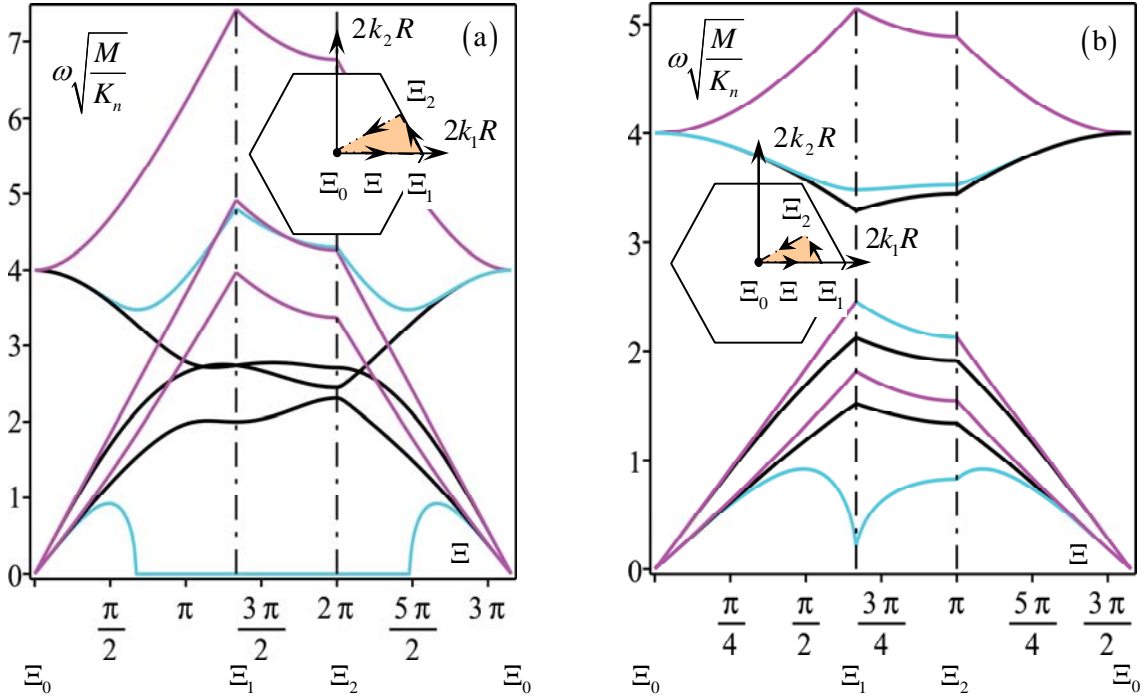

Figure 2: Non-dimensional frequency band structures for hexagonal packed granular material ($r = 1/50$, $r_1 = 2/3$). Comparison between the Lagrangian model (black line) and the homogenized models characterized by micropolar constitutive constant $\gamma$ (cyan line) and $\gamma^+$ (violet line), respectively. (a) Dispersive functions along the boundary of the non-dimensional irreducible first Brillouin zone; (b) Dispersive functions along the boundary of the sub-domain of the non-dimensional irreducible first Brillouin zone.

From Figure 2.a, referring to the realistic value $r_1 = 2/3$, i.e. $K_t = \frac{2}{3} K_n$, some features and drawbacks of the two micropolar models appear. The micropolar model derived through the standard continualization, despite its energetical inconsistency ($\gamma < 0$), exhibits a good performance in the long wavelength regime. It is worth to note the excellent accuracy in simulating the optical branch of the Lagrangian model with decreasing angular frequency, i.e. negative group velocity, for increasing the non-dimensional wave number. Conversely, the energetically consistent micropolar model ($\gamma^+ > 0$) exhibits an opposite behaviour, i.e. positive group velocity, that makes such model unreliable to simulate the optical acoustic branch of the Lagrangian model. This drawback have been observed and discussed by Merkel *et al.* (2011) and by Merkel and Luding (2017), also on the base of experimental results. At the same time, the energetical inconsistency of the micropolar model obtained by the standard continualization ($\gamma < 0$) has consequences on the acoustic branches when increasing the wave number (see Figure 2.a and 2.b) up to get vanishing values of the angular frequency, when the Legendre–Hadamard ellipticity conditions no longer holds. It must be remarked that for artificially low values of the ratio $r_1 = K_t/K_n$ are assumed as in the exemplary diagrams of Figure 3.a and 3.b, for which the parameter $\gamma > 0$, the qualitative difference in the optical branches between the two micropolar models disappears.

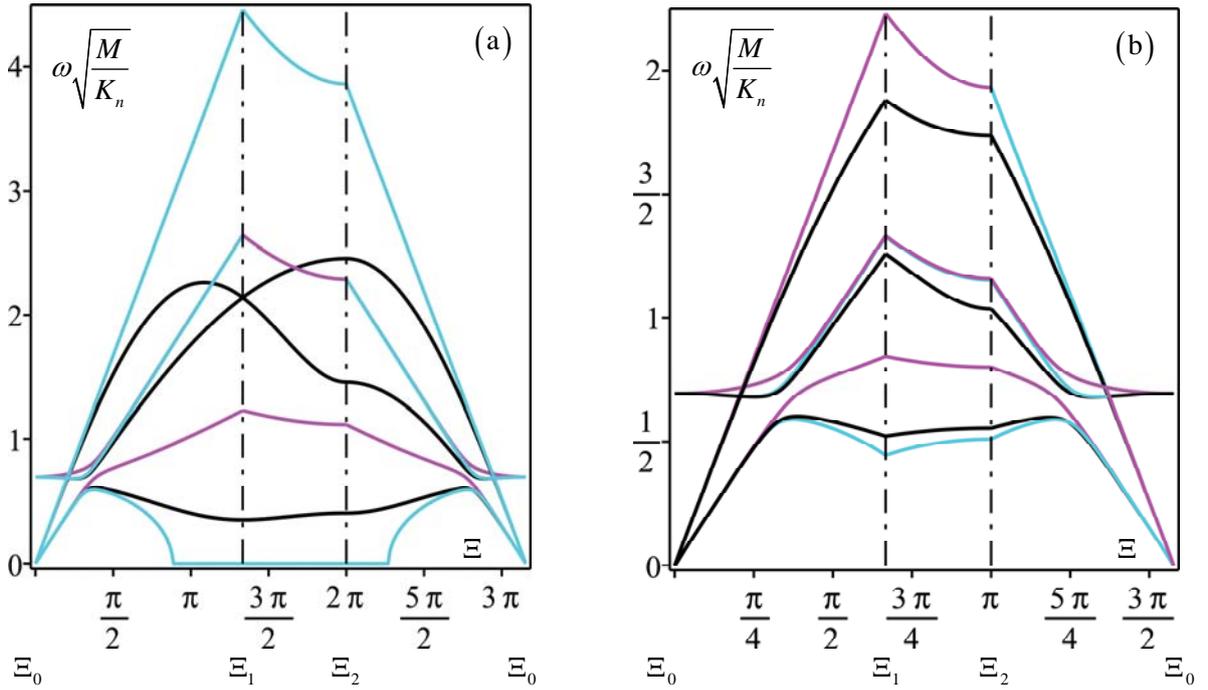

Figure 3: Frequency band structures for hexagonal packed granular material ($r = 1/50$, $r_1 = 1/50$). Comparison between the Lagrangian model (black line) and the homogenized models characterized by micropolar constitutive constant $S$ (cyan line) and $S^+$ (violet line), respectively. (a) Dispersive functions along the boundary of the non-dimensional irreducible first Brillouin zone; (b) Dispersive functions along the boundary of the sub-domain of the non-dimensional irreducible first Brillouin zone.

These considerations show that the continuous micropolar derived by the standard continualization presents the pathology of not being unconditionally positive defined (convex elastic potential energy) but it provides very good approximations of all the branches in the acoustic spectrum. Conversely, the energetically consistent micropolar model is not capable of providing qualitatively acceptable approximations of the optical branch. As neither approach provides physically acceptable solutions, it is considered appropriate to proceed with a different approach of homogenization.

## 4. Enhanced micropolar continuum

Let us reconsider the equation of motion (1) with the formalism of the shift operator, namely a linear transformation $\mathbf{d}_i = E_i \mathbf{d}$ correlating the displacement vector $\mathbf{d}_i(t) = \{u_i(t) \ v_i(t) \ \varphi_i(t)\}^T$ of the $i$-th circle to the displacement of the reference $\mathbf{d} = \{u(t) \ v(t) \ \varphi(t)\}^T$ of the reference circle (see Andrianov and Awrejcewicz, 2008; Bacigalupo and Gambarotta, 2019). In consideration of the hexagonal geometry of the material pattern, the shift operator may be written in the form $E_i(D_{\tilde{x}}^i) = \sum_{h=0}^{\infty} \frac{(2R)^h}{h!} (D_{\tilde{x}}^i)^h = \exp(2RD_{\tilde{x}}^i)$, with $D_{\tilde{x}}^i = \nabla \cdot \mathbf{n}^i = D_1 n_1^i + D_2 n_2^i = \frac{\partial}{\partial x_1} n_1^i + \frac{\partial}{\partial x_2} n_2^i$ and $\mathbf{n}_i$ the unit vector of the $i$-th circle. Accordingly, the motion of the circles surrounding the reference one are written in the form

$$\mathbf{d}_1(t) = E_1 \mathbf{d}(t), \quad \mathbf{d}_2(t) = E_2 \mathbf{d}(t), \quad \mathbf{d}_3(t) = E_3 \mathbf{d}(t),$$
$$\mathbf{d}_4(t) = E_4 \mathbf{d}(t), \quad \mathbf{d}_5(t) = E_5 \mathbf{d}(t), \quad \mathbf{d}_6(t) = E_6 \mathbf{d}(t). \tag{9}$$

with

$$E_1 = \exp\left[2RD_1\right], \quad E_2 = \exp\left[RD_1 + \sqrt{3}RD_2\right], \quad E_3 = \exp\left[-RD_1 + \sqrt{3}RD_2\right],$$
$$E_4 = \exp\left[-2RD_1\right], \quad E_5 = \exp\left[-RD_1 - \sqrt{3}RD_2\right], \quad E_6 = \exp\left[RD_1 - \sqrt{3}RD_2\right], \tag{10}$$

According to this notation the equation of motion (1) when ignoring the applied generalized forces may be rewritten as three pseudo-differential equations

$$\begin{cases} K_n \left[ E_4 - 3(r_1+1) + E_1 + \frac{1}{4}(3r_1+1)[E_2 + E_3 + E_5 + E_6] \right] u + \\ -\frac{\sqrt{3}}{4}(r_1-1) K_n [E_2 - E_3 + E_5 - E_6] v + \\ +\frac{\sqrt{3}}{2} r_1 K_n R [E_2 + E_3 - E_5 - E_6] \phi \end{cases} + M\ddot{u} = 0,$$

$$\begin{cases} -\frac{\sqrt{3}}{4}(r_1-1) K_n [E_2 - E_3 + E_5 - E_6] u + \\ +K_n \left[ r_1 E_1 - 3(r_1+1) + r_1 E_4 + \frac{1}{4}(3+r_1)[E_2 + E_3 + E_4 + E_6] \right] v + \\ -\frac{1}{2} r_1 K_n R [2E_1 + E_2 - E_3 - 2E_4 - E_5 + E_6] \phi \end{cases} + M\ddot{v} = 0,$$

$$\begin{cases} -\frac{\sqrt{3}}{2} r_1 K_n R [E_2 + E_3 - E_5 - E_6] u + \\ +\frac{1}{2} r_1 K_n R [2E_1 + E_2 - E_3 - 2E_4 - E_5 + E_6] v + \\ K_n R^2 \left\{ \frac{1}{12}(r^2 - 12 r_1)[E_1 + E_2 + E_3 + E_4 + E_5 + E_6] - \frac{1}{2}(r^2 + 12 r_1) \right\} \phi \end{cases} + J\ddot{\phi} = 0. \quad (11)$$

The homogenization procedure requires a downscaling law relating the displacement of the reference circle located at point **x**, namely the micro-displacement, to a macro-displacement field $\Upsilon(x_1, x_2, t) = \{U(x_1, x_2, t) \quad V(x_1, x_2, t) \quad \Phi(x_1, x_2, t)\}^T$, representing the macro-displacement and the macro-rotation, respectively. Here the scaling law is assumed according to Bacigalupo and Gambarotta (2021), which is a two-dimensional generalization of the first order regularization approach based on the central difference concept

$$\mathbf{d} = \left\{ \left( \prod_{j=1}^{n/2} D_{\tilde{x}_j} \right) \left[ \prod_{j=1}^{n/2} \left( \frac{\exp(2RD_{\tilde{x}_j}) - \exp(-2RD_{\tilde{x}_j})}{4R} \right) \right]^{-1} \Upsilon(x_1, x_2, t) \right\} \Bigg|_{\mathbf{x}}, \quad (12)$$

and for the hexagonal pattern takes the pseudodifferential form

$$\mathbf{d} = \left\{ \frac{16RD_1\left(RD_1 + \sqrt{3}RD_2\right)\left(-RD_1 + \sqrt{3}RD_2\right)}{[E_1 - E_4][E_2 - E_5][E_3 - E_6]} \right\} \Upsilon. \tag{13}$$

or in a synthetic form $\mathbf{d} = F(D_1, D_2)\Upsilon$, where the pseudo-function is deduced by equation (13). After substituting the scaling law (13) in equation (11) one obtains the pseudodifferential equation of motion for the equivalent continuum, i.e. in terms of the macro-displacement field

$$K_n \begin{bmatrix} P_{uu}(D_1, D_2) & P_{uv}(D_1, D_2) & P_{u\phi}(D_1, D_2) \\ P_{vu}(D_1, D_2) & P_{vv}(D_1, D_2) & P_{v\phi}(D_1, D_2) \\ P_{\phi u}(D_1, D_2) & P_{\phi v}(D_1, D_2) & P_{\phi\phi}(D_1, D_2) \end{bmatrix} \Upsilon - F(D_1, D_2) \begin{bmatrix} M & 0 & 0 \\ 0 & M & 0 \\ 0 & 0 & J \end{bmatrix} \ddot{\Upsilon} = \mathbf{0}, \tag{14}$$

where the pseudo-differential functions are defined as

$$P_{uu}(D_1, D_2) = \left[ E_4 - 3(r_1 + 1) + E_1 + \frac{1}{4}(3r_1 + 1)[E_2 + E_3 + E_5 + E_6] \right] F(D_1, D_2),$$

$$P_{uv}(D_1, D_2) = -\frac{\sqrt{3}}{4}(r_1 - 1)[E_2 - E_3 + E_5 - E_6] F(D_1, D_2),$$

$$P_{u\phi}(D_1, D_2) = \frac{\sqrt{3}}{2} r_1 R [E_2 + E_3 - E_5 - E_6] F(D_1, D_2),$$

$$P_{vv}(D_1, D_2) = \left[ r_1 E_1 - 3(r_1 + 1) + r_1 E_4 + \frac{1}{4}(3 + r_1)[E_2 + E_3 + E_5 + E_6] \right] F(D_1, D_2),$$

$$P_{v\phi}(D_1, D_2) = -\frac{1}{2} r_1 R [2E_1 + E_2 - E_3 - 2E_4 - E_5 + E_6] F(D_1, D_2),$$

$$P_{\phi\phi}(D_1, D_2) = R^2 \left\{ \frac{1}{12}(r^2 - 12r_1)[E_1 + E_2 + E_3 + E_4 + E_5 + E_6] - \frac{1}{2}(r^2 + 12r_1) \right\} F(D_1, D_2).$$

(15)

and the following symmetries hold $P_{vu}(D_1, D_2) = P_{uv}(D_1, D_2)$, $P_{\phi u}(D_1, D_2) = -P_{u\phi}(D_1, D_2)$, $P_{\phi v}(D_1, D_2) = -P_{v\phi}(D_1, D_2)$.

The pseudo-differential functions are expanded in formal Taylor series as follows

$$F(D_1, D_2) = 1 - R^2(D_1^2 + D_2^2) + \frac{48}{5}R^4(D_1^4 + 2D_1^2 D_2^2 + D_2^4) + \mathcal{O}(D^6),$$

$$P_{uu}(D_1, D_2) = \frac{3}{2}R^2\left[(r_1+3)D_1^2 + (3r_1+1)D_2^2\right] +$$
$$-\frac{R^4}{8}\left(11^4 r1 D_1 + 30 r1 D_1^2 D_2^2 + 27 r1 D_2^4 + 25 D_1^4 + 42 D_1^2 D_2^2 + 9 D_2^4\right) + \mathcal{O}(D^6),$$

$$P_{uv}(D_1, D_2) = -3(r_1-1)R^2 D_1 D_2 + \frac{1}{2}(r_1-1)R^4\left(5 D_1^3 D_2 + 3 D_1 D_2^3\right) + \mathcal{O}(D^6),$$

$$P_{u\phi}(D_1, D_2) = 6 r_1 R^2 D_2 - 3 r_1 R^4 D_2 (D_1^2 + D_2^2) + \mathcal{O}(D^6),$$

$$P_{vv}(D_1, D_2) = \frac{3}{2}R^2\left[(3r_1+1)D_1^2 + (r_1+3)D_2^2\right] +$$
$$-\frac{1}{8}R^4\left[(25 r_1 + 11)D_1^4 + 6(7 r_1 + 5)D_1^2 D_2^2 + 9(r_1 + 3)D_2^4\right] + \mathcal{O}(D^6),$$

$$P_{v\phi}(D_1, D_2) = -6 r_1 R^2 D_1 + 3 r_1 R^4 (D_1^3 + D_1 D_2^2) + \mathcal{O}(D^6),$$

$$P_{\phi\phi}(D_1, D_2) = -12 r_1 R^2 + \frac{1}{2}R^4 (12 r_1 + r^2)(D_1^2 + D_2^2) + \mathcal{O}(D^6),$$ (16)

where $D^\alpha(\cdot) = D_1^{\alpha_1} D_2^{\alpha_2}(\cdot) = \frac{\partial^{|\alpha|}}{\partial_{x_1}^{\alpha_1} \partial_{x_2}^{\alpha_2}}(\cdot)$, being $\alpha$ multi-index with length $|\alpha| = \alpha_1 + \alpha_2$ and $\alpha_1, \alpha_2 \in \mathbb{N}^*$.

By substituting in equation (14) the pseudo-differential functions obtained by (16) when retaining the terms up to the second order in the symbolic variable $D$, the differential equations of motion of the enhanced equivalent micropolar continuum are obtained

$$\begin{cases} \frac{\sqrt{3}}{8}K_n\left[(3+r_1)\frac{\partial^2 U}{\partial x_1^2} + (1+3r_1)\frac{\partial^2 U}{\partial x_2^2}\right] + \frac{\sqrt{3}}{4}(1-r_1)K_n\frac{\partial^2 V}{\partial x_1 \partial x_2} + \frac{\sqrt{3}}{2}r_1 K_n \frac{\partial \Phi}{\partial x_2} - \rho(\ddot{U} - R^2 \Delta \ddot{U}) = 0, \\ \frac{\sqrt{3}}{4}(1-r_1)K_n\frac{\partial^2 U}{\partial x_1 \partial x_2} + \frac{\sqrt{3}}{8}K_n\left[(1+3r_1)\frac{\partial^2 V}{\partial x_1^2} + (3+r_1)\frac{\partial^2 V}{\partial x_2^2}\right] - \frac{\sqrt{3}}{2}r_1 K_n \frac{\partial \Phi}{\partial x_1} - \rho(\ddot{V} - R^2 \Delta \ddot{V}) = 0, \\ \frac{\sqrt{3}}{24}K_n R^2 (r^2 + 12 r_1) \Delta \Phi - \sqrt{3} r_1 K_n \Phi + \frac{\sqrt{3}}{2}r_1 K_n \frac{\partial V}{\partial x_1} - \frac{\sqrt{3}}{2}r_1 K_n \frac{\partial U}{\partial x_2} - \frac{1}{2}\rho R^2 (\ddot{\Phi} - R^2 \Delta \ddot{\Phi}) = 0. \end{cases}$$ (17)

It is worth to note that such equations differs from equations (6) in *i*) the positive defined term $\frac{\sqrt{3}}{24}K_n R^2 (r^2 + 12 r_1) \Delta \Phi$ in the third equation and *ii*) in the presence of non-local inertia terms involving the Laplacian of the accelerations. This outcome makes the elasticity tensor in the constitutive equations

$$\begin{Bmatrix} \sigma_{11} \\ \sigma_{22} \\ \sigma_{12} \\ \sigma_{21} \\ m_1 \\ m_2 \end{Bmatrix} = \frac{\sqrt{3}}{8} K_n \begin{bmatrix} (3+r_1) & (1-r_1) & 0 & 0 & 0 & 0 \\ (1-r_1) & (3+r_1) & 0 & 0 & 0 & 0 \\ 0 & 0 & (1+3r_1) & (1-r_1) & 0 & 0 \\ 0 & 0 & (1-r_1) & (1+3r_1) & 0 & 0 \\ 0 & 0 & 0 & 0 & \frac{1}{3}R^2(r^2+12r_1) & 0 \\ 0 & 0 & 0 & 0 & 0 & \frac{1}{3}R^2(r^2+12r_1) \end{bmatrix} \begin{Bmatrix} \gamma_{11} \\ \gamma_{22} \\ \gamma_{12} \\ \gamma_{21} \\ \chi_1 \\ \chi_2 \end{Bmatrix} \quad (18)$$

to be unconditionally positive defined, being positive defined the elastic modulus relating micro-curvatures to micro, while the other three elastic moduli $\lambda$, $\mu$ and $\kappa$ are unchanged with respect to the micropolar standard model. The elastic potential energy density turns out to be positive defined as well as the kinetic energy density

$$T(\dot{U},\dot{V},\dot{\Phi}) = \frac{1}{2}\rho \begin{bmatrix} \dot{U}^2 + R^2\left(\frac{\partial \dot{U}}{\partial x_1}\right)^2 + R^2\left(\frac{\partial \dot{U}}{\partial x_2}\right)^2 + \dot{V}^2 + R^2\left(\frac{\partial \dot{V}}{\partial x_1}\right)^2 + R^2\left(\frac{\partial \dot{V}}{\partial x_2}\right)^2 + \\ +\frac{1}{2}R^2\dot{\Phi}^2 + \frac{1}{2}R^4\left(\frac{\partial \dot{\Phi}}{\partial x_1}\right)^2 + \frac{1}{2}R^4\left(\frac{\partial \dot{\Phi}}{\partial x_2}\right)^2 \end{bmatrix} \quad (19)$$

and the resulting micropolar model is confirmed to be energetically consistent.

The propagation of harmonic waves is governed by the eigenvalue problem

$$\begin{bmatrix} \frac{\sqrt{3}}{8}\begin{bmatrix}(3+r_1)k_1^2 + \\ +(1+3r_1)k_2^2\end{bmatrix} \\ -\frac{\rho}{K_n}\left[1+R^2(k_1^2+k_2^2)\right]\omega^2 & \frac{\sqrt{3}}{4}(1-r_1)k_1k_2 & -I\frac{\sqrt{3}}{2}r_1k_2 \\ \frac{\sqrt{3}}{4}(1-r_1)k_1k_2 & \frac{\sqrt{3}}{8}\begin{bmatrix}(1+3r_1)k_1^2 + \\ +(3+r_1)k_2^2\end{bmatrix} \\ -\frac{\rho}{K_n}\left[1+R^2(k_1^2+k_2^2)\right]\omega^2 & I\frac{\sqrt{3}}{2}r_1k_1 \\ I\frac{\sqrt{3}}{2}r_1k_2 & -I\frac{\sqrt{3}}{2}r_1k_1 & \frac{\sqrt{3}}{24}R^2(r^2+12r_1)(k_1^2+k_2^2)+\sqrt{3}r_1 + \\ -\frac{1}{2}\frac{\rho}{K_n}R^2\left[1+R^2(k_1^2+k_2^2)\right]\omega^2 \end{bmatrix} \begin{Bmatrix} \hat{U} \\ \hat{V} \\ \hat{\Phi} \end{Bmatrix} = \mathbf{0} \quad (20)$$

with two acoustic dispersion functions and one optical branch having al the long wavelength limit the same critical point of the Lagrangian model. Moreover it may be shown that the three dispersion functions approximate those by the Lagrangian model with good accuracy in the neighbourhood of the long wavelength limit, namely $\omega_{Lagrangian,i}(\mathbf{k}) = \omega_{enhanced,i}(\mathbf{k}) + \mathcal{O}(|\mathbf{k}|^3)$.

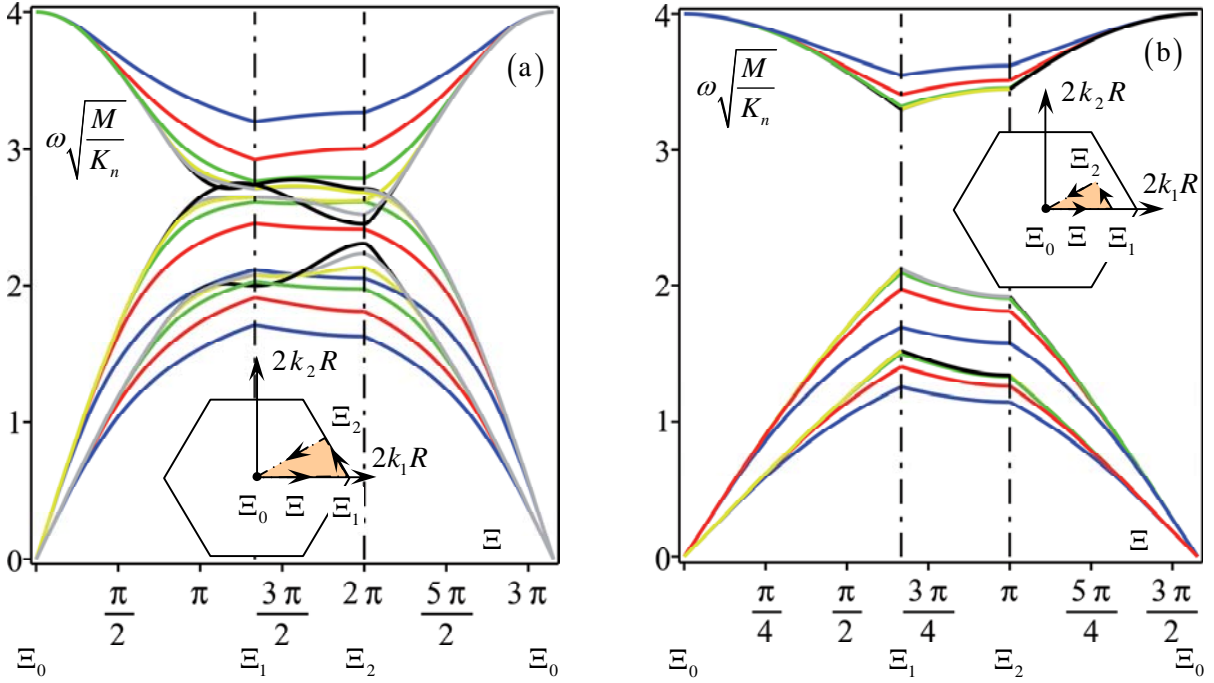

Figure 4: Frequency band structures for hexagonal packed granular material ($r = 1/50$, $r_1 = 2/3$). Comparison between the Lagrangian model (black line) and the homogenized models obtained via 2$^{nd}$ order (blue line), 4$^{th}$ order (red line), 8$^{th}$ order (green line) enhanced continualization, 16$^{th}$ order (yellow line) enhanced continualization and 32$^{nd}$ order (grey line) enhanced continualization. (a) Dispersive functions along the boundary of the non-dimensional irreducible first Brillouin zone; (b) Dispersive functions along the boundary of the sub-domain of the non-dimensional irreducible first Brillouin zone.

The same examples presented and discussed in Section 3 are here considered by comparing the dispersion functions of the Lagrangian model with the continuum models obtained by the enhanced micropolar homogenization here proposed. The results obtained by the eigenvalue problem (20) referred to the second order enhanced micropolar model are here complemented with higher order micropolar model obtained by retaining higher order terms in equations (16) up to the 32$^{nd}$ order. From this sequence of solutions the accuracy in the simulation and the convergence of the dispersion functions provided by the enhanced model to those of the Lagrangian model are shown, the latter result having being theoretically demonstrated in Bacigalupo and Gambarotta (2021). From the diagrams of Figure 4.a and 4.b the accuracy of the second order micropolar model is shown for $2R|\mathbf{k}| \leq \frac{\pi}{2}$. Higher order micropolar models appear to be more and more reliable.

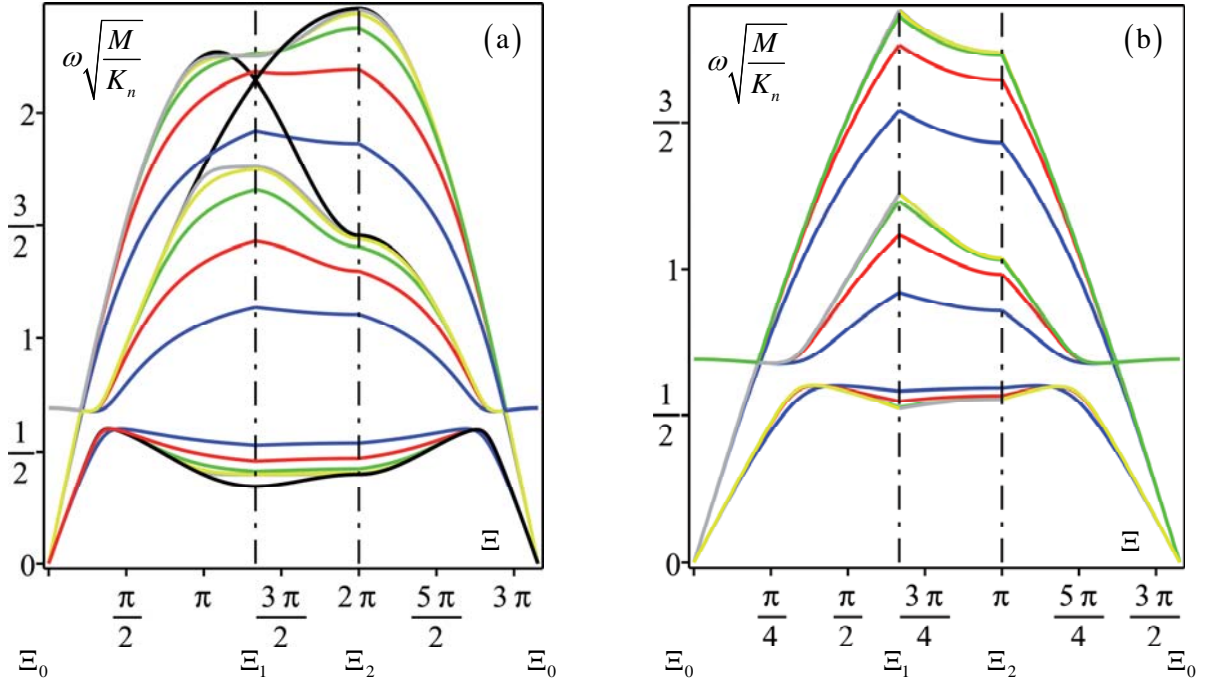

Figure 5: Frequency band structures for hexagonal packed granular material ($r = 1/50$, $r_1 = 1/50$). Comparison between the Lagrangian model (black line) and the homogenized models obtained via 2$^{nd}$ order (blue line), 4$^{th}$ order (red line), 8$^{th}$ order (green line) enhanced continualization, 16$^{th}$ order (yellow line) enhanced continualization and 32$^{nd}$ order (grey line) enhanced continualization. (a) Dispersive functions along the boundary of the non-dimensional irreducible first Brillouin zone; (b) Dispersive functions along the boundary of the sub-domain of the non-dimensional irreducible first Brillouin zone.

Analogous results are obtained with the limit value $K_t = K_n/50$ and the results are shown in Figure 5.a and 5.b. Specifically, in the diagrams of Figure 5.a the accuracy of the second order model appears as in the previous example for $2R|\mathbf{k}| \leq \frac{\pi}{2}$. Also a good convergence appears increasing the order of the micropolar model. It is worth to note that the inconsistency discussed in the conclusion of Section 3 has been solved. In fact, the elastic constitutive part of the micropolar model turns out to be energetically consistent, nevertheless the enhanced model, by virtue of the non-local inertia terms, results to be able to simulate with a good accuracy the curvature of the optical branch in the Lagrangian model, a peculiarity that is precluded by the classic micropolar model.

**6. Conclusions**

The problem about the inability of the micropolar model to simulate the optical band structure of hexagonally packed granular materials in the neighborhood of long wavelengths raised by Merkel et al. (2010, 2011) has been tackled in the present paper. In fact, the positive definiteness of the constitutive elastic tensor implies the optical branch in the frequency spectrum of the micropolar model to be characterized by an upward concavity in the neighborhood of the long wavelength regime while, on the contrary, the discrete granular model is characterized by a downward concavity of the optical branch. It has here shown that if a standard continualization of the difference equation of motion of the discrete model is carried out, an equivalent micropolar continuum is obtained which is able to accurately simulate the optical branches of the discrete model. Nevertheless, this homogenized continuum presents non-positive defined elastic potential energy, a deficiency that limits the reliability of the model and implies instability phenomena (*destabilizing effects*) in the acoustic branches.

To circumvent such drawbacks, an enhanced dynamic homogenization procedure for periodic granular materials has been proposed. The approach, which is based on an enhanced continualization via first order regularization together with a transformation of the governing difference equations of motion of the discrete system into pseudo-differential equations. Through a formal Taylor expansion of the pseudo-differential operators a higher order differential equation corresponding to the governing equation of a non-local continuum thermodynamically consistent has been obtained. The resulting approach allows obtaining an equivalent micropolar continuum characterized by inertial non-locality and higher order continua with non-local constitutive and inertial terms. It has been shown that these models are able to accurately describe both the static and dynamic behavior of the discrete granular model. Finally, the convergence to the response of the discrete system is shown when increasing the order of the higher order continuum.

Although this study is limited to the continuum modeling of a particular assembly of granules, however it is believed that this approach may constitute a contribution to the formulation of thermodynamically consistent micropolar continuum models and capable to provide accurate description of the frequency band structure of granular materials through enrichment of the constitutive model with inertial non-localities.

**Acknowledgements**

The authors acknowledge financial support from National Group of Mathematical Physics (GNFM-INdAM), from Compagnia San Paolo, project MINIERA no. I34I20000380007, and from University of Trento, project UNMASK 2020.

**Appendix A**

The equations of motion of a two dimensional micropolar continuum with hexagonal symmetry are (see for instance Eremeyev *et al.*, 2013; Nowacki, 1986; Wang and Stronge, 1999)

$$\begin{cases} (\lambda+2\mu)\dfrac{\partial^2 U}{\partial x_1^2} + (\mu+\kappa)\dfrac{\partial^2 U}{\partial x_2^2} + (\lambda+\mu-\kappa)\dfrac{\partial^2 V}{\partial x_1 \partial x_2} + 2\kappa\dfrac{\partial \Phi}{\partial x_2} - \rho\ddot{U} = 0, \\ (\lambda+\mu-\kappa)\dfrac{\partial^2 U}{\partial x_1 \partial x_2} + (\mu+\kappa)\dfrac{\partial^2 V}{\partial x_1^2} + (\lambda+2\mu)\dfrac{\partial^2 V}{\partial x_2^2} - 2\kappa\dfrac{\partial \Phi}{\partial x_1} - \rho\ddot{V} = 0, \\ \gamma\Delta\Phi - 4\kappa\Phi - 2\kappa\dfrac{\partial U}{\partial x_2} + 2\kappa\dfrac{\partial V}{\partial x_1} - j\ddot{\Phi} = 0, \end{cases}$$

(21)

with the four micropolar elastic moduli $\lambda$, $\mu$, $\kappa$ and $\gamma$. These moduli are subjected to the following condition to guarantee the positive definiteness of the elastic potential energy density $\lambda+\mu_n > 0$, $\mu>0$, $\kappa>0$ $\gamma>0$. These moduli appear in the constitutive equations

$$\begin{Bmatrix} \sigma_{11} \\ \sigma_{22} \\ \sigma_{12} \\ \sigma_{21} \\ m_1 \\ m_2 \end{Bmatrix} = \begin{bmatrix} 2\mu+\lambda & \lambda & 0 & 0 & 0 & 0 \\ \lambda & 2\mu+\lambda & 0 & 0 & 0 & 0 \\ 0 & 0 & \mu+k & \mu-k & 0 & 0 \\ 0 & 0 & \mu-k & \mu+k & 0 & 0 \\ 0 & 0 & 0 & 0 & \gamma & 0 \\ 0 & 0 & 0 & 0 & 0 & \gamma \end{bmatrix} \begin{Bmatrix} \gamma_{11} \\ \gamma_{22} \\ \gamma_{12} \\ \gamma_{21} \\ \chi_1 \\ \chi_2 \end{Bmatrix}, \quad (22)$$

involving the asymmetric stress tensor components $\sigma_{11}, \sigma_{12}, \sigma_{21}, \sigma_{22}$, the couple stress components $m_1$ and $m_2$, the asymmetric strain tensor components $\gamma_{11} = u_{1,1}$, $\gamma_{22} = u_{2,2}$, $\gamma_{12} = u_{1,2} + \phi$, $\gamma_{21} = u_{2,1} - \phi$ and the curvature components $\chi_1 = \phi_{,1}$ and $\chi_2 = \phi_{,2}$.

Finally, the harmonic wave propagation is governed by the eigenvalue problem

$$\begin{bmatrix} (2\mu+\lambda)k_1^2 + (\mu+\kappa)k_2^2 - \rho\omega^2 & (\mu+\lambda-\kappa)k_1 k_2 & -2i\kappa k_2 \\ (\mu+\lambda-\kappa)k_1 k_2 & (\mu+\kappa)k_1^2 + (2\mu+\lambda)k_2^2 - \rho\omega^2 & 2i\kappa k_1 \\ 2i\kappa k_2 & -2i\kappa k_1 & \gamma(k_1^2 + k_2^2) + 4\kappa - I\omega^2 \end{bmatrix} \begin{Bmatrix} \hat{U} \\ \hat{V} \\ \hat{\Phi} \end{Bmatrix} = \mathbf{0}$$

(23)

with $\omega$ angular frequency, $\hat{\mathbf{d}} = \{\hat{U} \quad \hat{V} \quad \hat{\Phi}\}^T$ polarization vector, $k_1$ and $k_2$ components of the wave vector.